\documentclass[preprintnumbers,secnumarabic,amsmath,amssymb,nofootinbib,floatfix,superscriptaddress,english,12pt]{article}
\usepackage{array}
\usepackage{graphicx}
\usepackage{amssymb}
\usepackage[english]{babel}
\usepackage{amsmath}
\usepackage{multirow}
\usepackage{prettyref}
\usepackage{babel}
\usepackage{units}
\usepackage[latin1]{inputenc}
\usepackage{amsfonts}
\usepackage{amssymb}
\usepackage{babel}
\usepackage{color}
\usepackage{cite}
\def\@fmsl@sh#1#2#3{\m@th\ooalign{$\hfil#1\mkern#2/\hfil$\crcr$#1#3$}}
 \def\eq#1\en{\begin{equation}#1\end{equation}}
\def\s[#1,#2]{[#1\stackrel{\star}{,}#2]}
\def\sx[#1,#2]{[#1\stackrel{\star_{x}}{,}#2]}

\newcommand{\non}{\nonumber}
\newcommand{\ba}{\begin{eqnarray}}
\newcommand{\ea}{\end{eqnarray}}
\newcommand{\D}{\nabla}

\newcommand{\nc}{\newcommand}
\nc{\beq}{\begin{equation}}
\nc{\eeq}{\end{equation}}
\nc{\beqa}{\begin{eqnarray}}
\nc{\eeqa}{\end{eqnarray}}

\def\bc{\begin{center}}
\def\ec{\end{center}}

\def\to{\rightarrow}

\def\gsim{\mathrel{\mathpalette\atversim>}}

\def\bc{\begin{center}}
\def\ec{\end{center}}

\def\gsim{\mathrel{\rlap{\lower4pt\hbox{\hskip1pt$\sim$}}

    \raise1pt\hbox{$>$}}}       

\def\gsim{\mathrel{\rlap{\lower4pt\hbox{\hskip1pt$\sim$}}
    \raise1pt\hbox{$>$}}}       

\begin{document}

\makeatletter
\def\fmslash{\@ifnextchar[{\fmsl@sh}{\fmsl@sh[0mu]}}
\def\fmsl@sh[#1]#2{%
  \mathchoice
    {\@fmsl@sh\displaystyle{#1}{#2}}%
    {\@fmsl@sh\textstyle{#1}{#2}}%
    {\@fmsl@sh\scriptstyle{#1}{#2}}%
    {\@fmsl@sh\scriptscriptstyle{#1}{#2}}}
\def\@fmsl@sh#1#2#3{\m@th\ooalign{$\hfil#1\mkern#2/\hfil$\crcr$#1#3$}}
\makeatother

\thispagestyle{empty}
\begin{titlepage}
\boldmath
\begin{center}
  \Large {\bf    Imprints of Quantum Gravity in the Cosmic Microwave Background}
    \end{center}
\unboldmath
\vspace{0.2cm}
\begin{center}
{  {\large Xavier Calmet}$^a$\footnote{x.calmet@sussex.ac.uk}},
{  {\large James Edholm}$^b$\footnote{j.edholm@lancaster.ac.uk}},
{and}
{  {\large Iber\^e Kuntz}$^a$\footnote{ibere.kuntz@sussex.ac.uk}}
 \end{center}
\begin{center}
{\sl $^a$Department of Physics and Astronomy, 
University of Sussex, Brighton, BN1 9QH, United Kingdom
}
\\
{\sl $^b$Lancaster University, Lancaster, LA1 4YW, United Kingdom
}

\end{center}
\vspace{5cm}
\begin{abstract}
\noindent
It has been shown that the spectrum of quantum gravity contains at least two new modes in addition to the massless graviton: a massive spin-0 and a massive spin-2. We calculate their power spectrum during inflation and we argue that they could leave an imprint in the cosmic microwave background should their masses be below the inflationary scale.
\end{abstract}  
\vspace{5cm}
\end{titlepage}



\newpage

 \section{Introduction}
 
A fully consistent theory of quantum gravity remains an elusive dream. While there are promising ideas on how to build a consistent quantum theory of gravity such as string theory or loop quantum gravity, there is no consensus yet on the right theoretical approach to the problem. The only strong constraint that one can impose is that classical general relativity must emerge as a low energy  effective action of any viable quantum theory of gravity. If one is interested in physics at energies below the Planck scale or some $10^{19}$ GeV, it turns out that this constraint is enough to enable us to make predictions in quantum gravity independently of the ultraviolet theory. The framework that enables such predictions is that of effective field theory.

The application  of effective field theory  methods to quantum gravity  \cite{Weinberg,Bar1984,Bar1985,Buchbinder:1992rb,Donoghue:1994dn,Calmet:2013hfa} is a powerful approach for making low-energy quantum predictions without needing detailed knowledge of the ultraviolet physics.  It has the advantage of separating out the ultraviolet from the infrared physics, opening up the possibility for model-independent predictions in the latter while keeping the ultraviolet nature of the theory encoded in the Wilson coefficients. Thus, in principle, it gives us a way to probe quantum gravity experimentally up to the Planck scale. At this energy scale, contributions from an infinite number of curvature invariants would become important, signaling the breakdown of the effective theory approach. 

A key prediction of this effective field theory approach is that there are more degrees of freedom than just the massless graviton in quantum gravity \cite{Calmet:2014gya,Calmet:2016sba,Stelle:1977ry,Calmet:2017omb,Calmet:2018qwg}. Indeed, there are at least two new massive fields with spin-0 and spin-2. In this paper we propose to search for the imprints of these new bosons in the cosmic microwave background as they could be excited during inflation if the scale of inflation larger than the masses of these fields. Indeed, any light gravitational field can be excited during inflation as these fields couple to the energy-momentum tensor like the metric does. 
This is true for spin-0 and spin-2 bosons. Our main result is a calculation of the power spectrum 
associated with these massive modes in quantum gravity.  Experiments on earth, and the E\"ot-Wash 
experiment in particular only set very weak bounds on the masses of these new modes. These masses must be larger than $1 \times 10^{-12}$ GeV \cite{Calmet:2018uub}, otherwise a modification of Newton's $1/r$ law would have been observed by torsion pendulum experiments. Depending on their masses and the scale of inflation, these new modes could have been excited in full analogy to the gravitational wave power spectrum expected from the excitation of the massless graviton mode.

The local part of the effective action for quantum gravity is given by
\ba \label{eq:localaction}
        S= \frac{M_p^2}{2}\int d^4x \sqrt{-\tilde g} \left(\tilde R + b_1 \tilde R^2 
        + b_2 \tilde R_{\mu\nu} \tilde R^{\mu\nu} + b_3 
        \tilde R_{\mu\nu\rho\sigma}\tilde R^{\mu\nu\rho\sigma}\right),
\ea
where $\tilde R$, $\tilde R_{\mu\nu}$ and $\tilde R_{\mu\nu\rho\sigma}$ are the Ricci scalar, Ricci tensor and Riemann tensor of the metric $\tilde g_{\mu\nu}$ respectively. It is straightforward to see that this action contains three modes: a massless spin-2 that corresponds to the graviton, a massive spin-2 and a massive spin-0.  The local part of the effective quantum gravitation is identical to the extension of general relativity studied by Stelle long ago \cite{Stelle:1977ry}.  These new fields can be made explicit in the action through a field redefinition $\tilde g_{\mu\nu}\to g_{\mu\nu}$, in which case the action reads \cite{Hindawi:1995an,Stelle:1977ry}
\ba
        S&=&\frac{M_p^2}{2}\int \sqrt{-g} \bigg[R -\frac{3}{2} \left(
        A^{-1}(\phi_{\sigma\tau})\right)_\mu{}^\nu \nabla^\mu \chi \nabla_\nu 
        \chi -\frac{3}{2} m_0^2\left(\det A(\phi_{\sigma\tau})\right)^{-1/2}
        (1-e^{-\chi})^2 \non\\
        &&-g^{\mu\nu} \left(C^\lambda{}_{\mu\rho} 
        (\phi_{\sigma\tau}) C^\rho{}_{\nu\lambda} (\phi_{\sigma\tau}) - 
        C^\lambda{}_{\mu\nu} (\phi_{\sigma\tau}) C^\rho_{\rho\lambda}(\phi_{\sigma\tau})\right)
        \non \\ &&
        +\frac{1}{4}m_2^2 \left(\det A(\phi_{\sigma\tau})\right)^{-1/2}
        \left(\phi_{\mu\nu} \phi^{\mu\nu}-\phi^2\right)\bigg],
\ea      
where $m_0^{-2} = 6b_1 + 2b_2 + 2b_3$ is the mass of the scalar field $\chi$ and $m_2^{-2}=-b_2-4b_3$ is the mass of the massive spin-2 particle $\phi_{\mu\nu}$. 
$A_{\nu}{}^\lambda$ is defined as 
\ba
        A_{\nu}{}^\lambda = \left(1+\frac{1}{2}\phi\right)\delta_{\nu}{}^\lambda - \phi_{\nu}{}^\lambda,
\ea        
and 
\ba
        C^\lambda{}_{\mu\nu} = \frac{1}{2} \left(\tilde g^{-1}\right)^{\lambda\rho}
        \left({\D}_\mu \tilde g_{\nu\rho} + {\D}_\nu \tilde g_{\mu\rho} - {\D}_\rho \tilde g_{\mu\nu}\right),
\ea        
where the connection $\D$ is compatible with the new metric $g_{\mu\nu}$.
\section{Linearisation around de Sitter}
Upon linearisation around a background de Sitter spacetime $\bar{g}_{\mu\nu}$, i.e. $g_{\mu\nu} = \bar g_{\mu\nu} + h_{\mu\nu}$, this action simplifies to
\ba
S = \frac{M_p^2}{2}\int\sqrt{-g}[\bar R + \mathcal{L}_\text{g} + \mathcal{L}_s + \mathcal{L}_\text{PF}],
\ea
where $\bar R$ is the background curvature,
\ba
        \mathcal{L}_\text{g} = -\frac{1}{4} \left(\nabla_\mu h \nabla^\mu h - \nabla_\mu h_{\nu\rho} \nabla^\mu h^{\nu\rho}+2 \nabla^\mu h_{\mu\nu} \D^\nu h - 2 \D_\mu h_{\nu\rho}
         \D^\rho h^{\mu\nu}\right),
\ea        
is the graviton sector,
\ba
        \mathcal{L}_s = -\frac{3}{2} \bar{\nabla}^\mu \chi \bar{\nabla}_\mu \chi 
        -\frac{3}{2} m_0^2(1-e^{-\chi})^2,
\label{eq:s}
\ea
is the scalar sector and
\ba
\mathcal{L}_\text{PF} = -\frac{1}{4} \left(\nabla_\mu \phi \nabla^\mu \phi 
-\nabla_\mu \phi_{\nu\rho} \nabla^\mu \phi^{\nu\rho}
+2 \nabla^\mu \phi_{\mu\nu} \D^\nu \phi - 2 \D_\mu \phi_{\nu\rho}\D^\rho \phi^{\mu\nu}\right) 
+ \frac{1}{4} m^2_2 \left(\phi_{\mu\nu} \phi^{\mu\nu} -\phi^2\right), 
\label{eq:pf}
\ea
is the massive spin-2 sector. Note that \eqref{eq:s} is the same Lagrangian of the scalaron of Starobinsky gravity. Its curvature power spectrum is already known \cite{Guzzetti:2016mkm}
\ba
\Delta_\zeta^2 = \frac{1}{8\pi^2\epsilon}\frac{H^2}{M_p^2},
\label{eq:psc}
\ea
where $\epsilon$ is the slow-roll parameter of the inflation model being considered. The power spectrum of the graviton is also known  \cite{Baumann:2009ds}
\ba
\Delta^2_g = \frac{2H^2}{\pi^2 M_p^2}.
\ea
In contrast to the curvature power spectrum, the graviton power spectrum is model independent. This is a well established fact.

\section{Power spectrum of massive gravitational waves}
We are then left with the job of calculating the power spectrum of massive gravitational waves described by \eqref{eq:pf}. Gravitational waves are represented by tensor perturbations $\phi_{ij}$, which satisfy $\partial_i \phi^{ij} = \phi^i_i = 0$. For a de Sitter background in conformal coordinates, i.e.
\ba
ds^2 = a(\tau)^2(-d\tau^2 + dx_i dx^i),
\ea
the action of the massive spin-2 becomes
\ba
S_\text{PF} = \frac{M_p^2}{2}\int\mathrm{d}^4x \frac{a^2}{4}\left[-\phi'^2_{ij} + (\nabla\phi_{ij})^2 + a^2 m_2^2 \phi_{ij}^2 \right],
\ea
where prime represents differentiation with respect to the conformal time. We can canonically normalize this action by making the following field redefinition:
\ba
v_{ij} = \frac{a}{2}M_p \phi_{ij}.
\ea
The equation of motion for $v_{ij}$ is then
\ba
v''_{ij} - \nabla^2 v_{ij} + m^2 v_{ij}^2 = 0,
\label{eq:eom}
\ea
where
\ba
m^2 \equiv a^2m_2^2 - \frac{a''}{a} = \left(\frac{m_2^2}{H^2}-2\right)\frac{1}{\tau^2}.
\ea
By solving \eqref{eq:eom} in the Fourier space at superhorizon scales with the Bunch-Davies initial condition, we find the power spectrum of $v_{ij}$:
\ba
P_v = |v_{k,\gamma}|^2 = \frac{|\tau|}{4\pi}\Gamma^2(n)\left(\frac{2}{k|\tau|}\right)^{2n},
\ea
where $\Gamma$ is the gamma function and $n=\sqrt{\tfrac{9}{4}-\tfrac{m_2^2}{H^2}}$. The power spectrum of the massive spin-2 field $\phi_{ij}$ is then
\begin{align} \label{eq:powerspectrum1}
\Delta^2_\text{PF} &= 2\left(\frac{k^3}{2\pi^2}\right)\left(\frac{2}{aM_p}\right)^2P_v\\
\label{eq:powerspectrum2}
&= \frac{2^{2n}\Gamma^2(n)k^{3-2n}}{\pi^3 a^2 M_p^2} (aH)^{2n-1}.
\end{align}
The multiplicative factor 2 in \eqref{eq:powerspectrum1} accounts for two of five polarizations of the massive spin-2 mode. The remaining polarizations contribute to a vector power spectrum which is however not observable.  If we assume $m_2\ll H$, then 
$n \approx \frac{3}{2} -\frac{1}{3} \frac{m_2^2}{H^2},$       
and Eq. \eqref{eq:powerspectrum2} further simplifies to
\ba
\Delta^2_\text{PF} = \frac{2H^2}{\pi^2 M_p^2}\left(\frac{k}{aH}\right)^{\frac{2m_2^2}{3H^2}}.
\ea
Thus, the mass of the spin-2 particle has little influence on the amplitude $A_\text{PF} = \frac{2H^2}{\pi^2 M_p^2} \sim 3.5\times 10^{-14}$, where we used $H\sim 10^{12}$GeV, but it has a measurable effect on the tensor tilt:
\ba
n_\text{PF} &=& \frac{d\ln\Delta^2_\text{PF}}{d\ln k} = \frac{2m_2^2}{3H^2}.
\ea
In particular, the massive spin-2 particle produces a blue tilt as $n_\text{PF}$ is positive. In a quasi-de Sitter background, the tensor tilt would become $n_\text{PF} = -2\epsilon + \tfrac{2m_2^2}{3H^2}$, so the magnitude of $m_2^2/H^2$ would have to overcome the slow-roll parameter $\epsilon$ in order to produce a blue tilt.

\section{Inflationary observables}
The total tensor power spectrum ultimately depends on the power transmitted to the CMB photons. Therefore, the coupling of the tensor fields to matter dictates how their power spectra are combined \cite{Fasiello:2015csa}. The matter action reads \cite{Calmet:2018qwg}
\ba
S_m = \int\mathrm{d}^4x\sqrt{-\bar g}(h_{\mu\nu}-k_{\mu\nu})T^{\mu\nu},
\ea 
where $T^{\mu\nu}$ is the energy-momentum tensor. The net tensor spectrum is then
\ba
\Delta_t^2 = \Delta_g^2 + \Delta_\text{PF}^2 = \frac{2H^2}{\pi^2 M_p^2}\left(\frac{k}{aH}\right)^{-2\epsilon}\left[1 + \left(\frac{k}{aH}\right)^{\frac{2m_2^2}{3H^2}}\right].
\ea
Consequently, the net tensor tilt is
\ba
n_t = -4\epsilon + \frac{2m_2^2}{3H^2},
\label{eq:nt}
\ea
and the tensor-to-scalar ratio evaluated at some pivot scale $k_*$ becomes
\ba
r = \frac{\Delta^2_t(k_*)}{\Delta^2_\zeta(k_*)} = 32\epsilon.
\label{eq:r}
\ea
Note that the ratio $r$ is twice the standard value because of the contribution of the massive spin-2.

\begin{figure}[!htbp]
\centering
\includegraphics[scale=0.6]{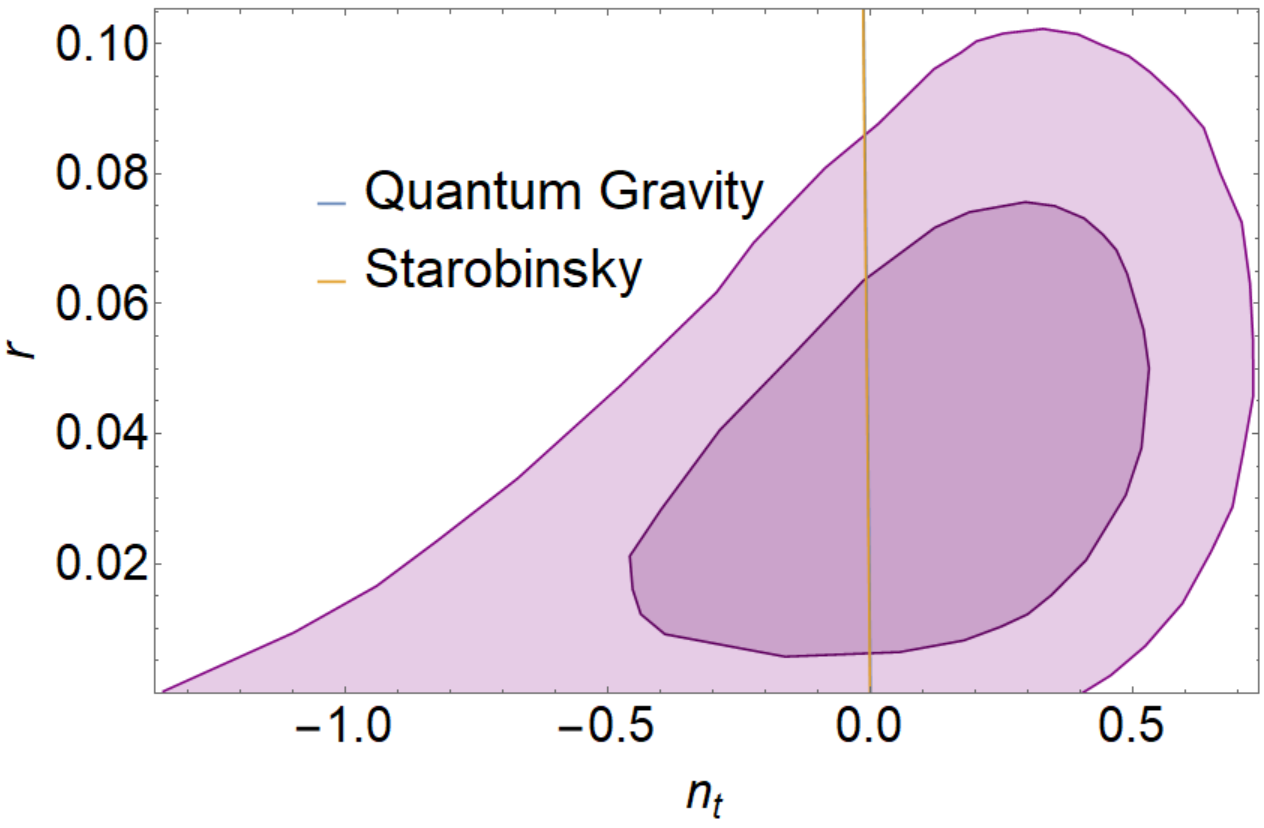}
\caption{We plot the tensor-scalar ratio $r$ against the tensor spectral tilt 
$n_t$ for Starobinsky inflation and Effective Quantum Gravity, using \eqref{eq:newconsistency} 
and choosing $m_2/H=0.001$, against the 2018 Planck results \cite{Akrami:2018odb}. The line for Starobinsky's model (orange line) overlaps the blue line corresponding to Quantum Gravity precisely, making it impossible to distinguish between both lines with current data.}
\label{ntfigure}
\end{figure}

From \eqref{eq:nt} and \eqref{eq:r}, we can see that the consistency condition for the tensor tilt is modified:
\ba \label{eq:newconsistency}
n_t = -\frac{r}{8} + \frac{2m_2^2}{3H^2}.
\ea
The increasing constraining power of B-mode polarization data allowed Planck collaboration to set bounds on $r$ without assuming the consistency condition. At $k_*=0.01\,\text{Mpc}^{-1}$, which corresponds to the decorrelation scale of $r$ and $n_t$, they found $r_{0.01}<0.091$ and $-0.34<n_t<2.63$ \cite{Akrami:2018odb}. The upper bound translates into
\ba
\frac{m_2}{H}\lesssim 1.99.
\label{eq:bound1}
\ea
We emphasize that this bound only makes sense when $m_2/H\ll 1$, in which case the massive spin-2 can be produced during inflation. Thus, Eq. \eqref{eq:bound1} is trivially satisfied and we conclude that current data cannot constrain $m_2$. In Fig. \ref{ntfigure}, we plot the 
tensor-scalar ratio against the tensor tilt, according to Eq. \eqref{eq:newconsistency}, to illustrate that quantum gravity
gives almost identical results to Starobinsky inflation given the  current observational bounds.

A blue spectrum at the CMB would be compatible with the existence of the massive spin-2 particle whose mass would be bounded by the experimental constraint of $n_t$. The observation of a red tilt, on the other hand, would imply
\ba
\frac{m_2}{H}<\sqrt{6\epsilon},
\ea
or using the bound $\epsilon<4.5\times 10^{-3}$ and $H\sim 10^{13}$GeV \cite{Akrami:2018odb},
\ba
m_2<1.6\times 10^{12}\text{GeV}\quad \text{(red power spectrum)}.
\ea
Clearly, this bound depends on the precise value of $H$ and only holds if the tensor tilt turns out to be red. The only possible way to solve for $m_2$ without using any estimation for $H$ is through a direct measurement of the tensor power spectrum. This would naturally lead to separate bounds on $H$ and $m_2$. Quantum gravity thus predict a blue tilt spectrum if the mass of the massive spin-2 field is bellow the scale of inflation with essentially no correction to the gravitational wave power spectrum. 

\begin{figure}[!htbp]
\centering
\includegraphics[scale=0.65]{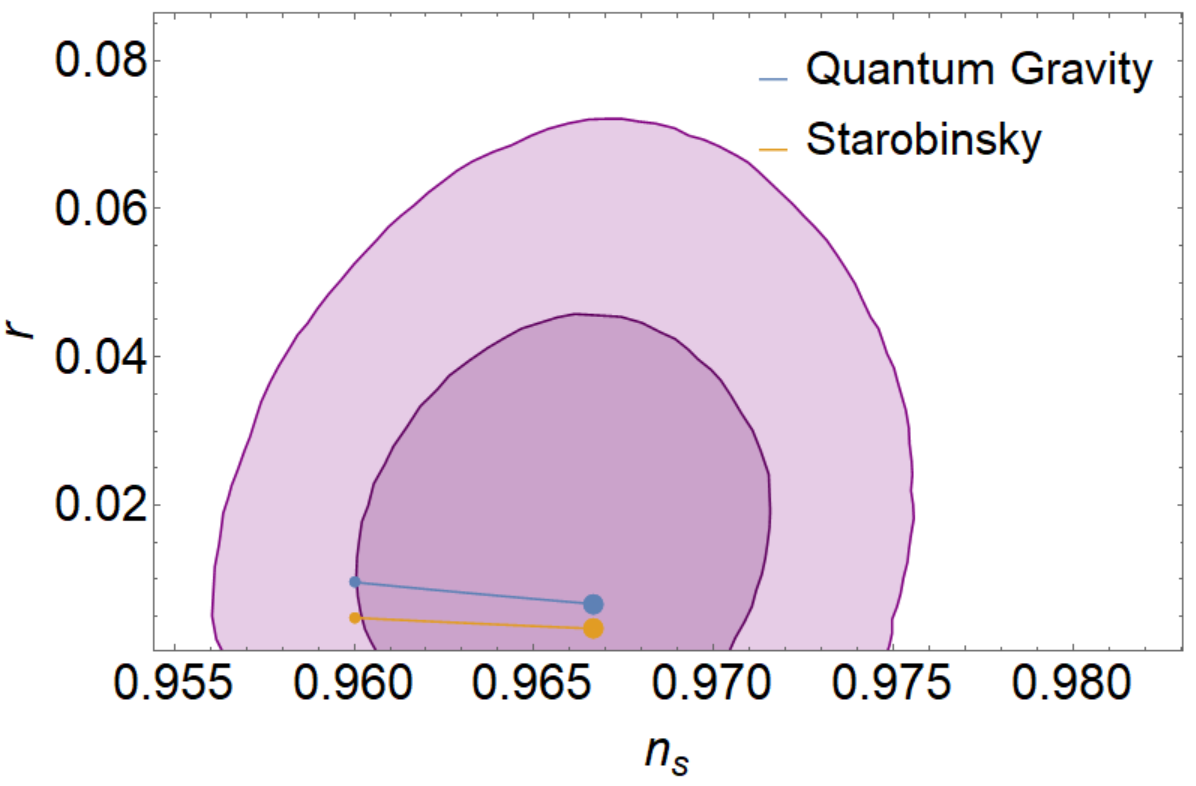}
\caption{We plot the tensor-scalar ratio $r$ against the scalar spectral tilt $n_s$ for Starobinsky inflation and Effective Quantum Gravity, 
using \eqref{eq:psc} and \eqref{eq:r}, against the 2018 Planck results \cite{Akrami:2018odb}. 
The small dot represents $N=50$ e-folds and the large dot represents $N=60$.}
\label{nsfigure}
\end{figure}

We must stress that we have not said anything about the nature of the inflaton so far, thus the above findings are somewhat model-independent. There are three possible scenarios depending on whether inflation is driven by: 1) the scalaron from $R^2$ due to quantum gravity, 2) an external scalar field or 3) both (e.g. multi-field inflation). In each case, we would have different predictions for the slow-roll parameter $\epsilon$, which would affect all observables. In the first case, for example, the slow-roll parameter can be written in terms of the number of e-foldings $N$ \cite{DeFelice:2010aj}:
\ba
\epsilon = \frac{3}{4 N^2},
\ea
and one finds
\ba
\label{eq:staro}
\label{eq:rscalaron}
r &=& \frac{24}{N^2},\\
\label{eq:nsscalaron}
n_s &=& 1 - \frac{2}{N},\\
n_t &=& -\frac{3}{N^2} + \frac{2m_2^2}{3H^2},
\ea
where $n_s$ is the spectral tilt. The additional multiplicative factor of 2 found in \eqref{eq:r} is carried over to \eqref{eq:staro}. We plot the tensor-scalar ratio \eqref{eq:rscalaron} against the spectral tilt \eqref{eq:nsscalaron} compared to the data from the 2018 Planck results in Fig. \ref{nsfigure}.

\section{Conclusion}
In this short paper, we have shown that quantum gravity could leave an imprint in the cosmic microwave background in the form of massive gravitational waves. This is a model independent prediction which does not dependent on the ultra-violet completion of quantum gravity. Besides the power spectra of the inflaton and the usual gravitational waves (massless spin-2), quantum gravity predicts power spectra for the massive spin-2 and spin-0 modes present in the effective action. It is interesting that the massive spin-0 could actually be the inflaton itself, as it is the case in Starobinsky's inflationary model. Interplays between the scalaron and the Higgs field could also be of interest and have been studied in \cite{Calmet:2016fsr}.

If the masses of the massive spin-2 and spin-0 waves are below the scale of inflation, CMB observations have the potential to extend the search for quantum gravity massively. Currently, the bounds from the E\"ot-Wash torsion pendulum experiment simply imply that their masses are larger than $1 \times 10^{-12}$ GeV. The discovery of the power spectrum for gravitational waves would fix the scale of inflation and thereby enable one to set a limit on the masses of these new modes or with a bit of luck prelude to a discovery of the new states predicted by quantum gravity. Future detectors could find the value of $r$ to within $\pm0.001$ \cite{Matsumura:2014fte,Bouchet:2015arn}, meaning that we would have strong evidence either for or against lighter than the scale of inflation quantum gravitational fields.

We conclude by noting that the massive gravitational modes discussed here could also represent a large component of dark matter as discussed in \cite{Calmet:2018uub}. As explained in \cite{Calmet:2018uub}, these modes would then have to be very light and would essentially be massless during inflation. If these massive gravitational modes are truly the missing dark matter, then we expect that they will leave an imprint in the CMB as discussed here.

\bigskip{}

{\it Acknowledgments:}
This work supported in part  by the Science and Technology Facilities Council (grant number ST/P000819/1) and by the National Council for Scientific and Technological Development (CNPq - Brazil).


\end{document}